\documentclass[prl,twocolumn,floatfix]{revtex4-1}
\usepackage{graphicx}
\usepackage{times}
\begin{document}
\title{Shock-driven hydrodynamic instability induced by particle seeding}
\author{Peter Vorobieff, Michael Anderson, Joseph Conroy, Ross White, C. Randall Truman}
\affiliation{Department of Mechanical Engineering, The University of New Mexico, NM 87131, USA}
\author{Sanjay Kumar}
\affiliation{Department of Engineering, University of Texas - Brownsville, TX 78520, USA}
\begin{abstract}
We report an experimental observation of an instability in gas of constant density (air) with an initial non-uniform seeding of small droplets that develops as a planar shock wave passes through the gas-droplet mix. The seeding non-uniformity is produced by vertical injection of a slow-moving jet of air pre-mixed with glycol droplets into the test section of a shock tube, with the plane of the shock parallel to the axis of the jet. After the shock passage, we observe development of two counter-rotating vortices in the plane normal to that axis. The  physical mechanism of the instability we observe is peculiar to multiphase flow, where the shock acceleration causes the second (embedded) phase to move with respect to the embedding medium. With sufficient seeding concentration, this leads to entrainment of the embedding phase that acquires a relative velocity dependent on the initial seeding, resulting in vortex formation in 
the flow. 
\end{abstract}
\maketitle

Richtmyer-Meshkov instability (RMI) \cite{richtmyer,meshkov} develops on an initially perturbed interface between two gases of different densities subjected to an impulsive acceleration (shock).  Misalignment between the pressure gradient on the shock front 
and density gradients in the medium the shock passes through leads to vorticity deposition. 
After a short stage of linear perturbation growth described by the theory of Richtmyer \cite{richtmyer}, the flow driven by shock-deposited vortices enters a stage of 
nonlinear instability growth and eventually transitions to turbulence. RMI plays an important role in transient processes in Earth's magnetosphere \cite{geophys}, astrophysical phenomena such as evolution of supernova remnants \cite{astro1,astro2}, high energy density physics, specifically inertial confinement fusion (ICF) \cite{hedp}, and supersonic combustion 
\cite{scram}. 

In many of these processes, the shock actually propagates through a multiphase medium comprised of embedding gaseous phase and embedded droplets or particles. For example, in astrophysical processes, dusty plasma \cite{dusty} is ubiquitous. While the relative velocity of the embedded phase can be zero prior to the shock acceleration, the high density difference between the phases leads to the particles and droplets ``lagging behind'' the gas after the shock passage. This lag has to be taken into consideration in experiments where particles are used to trace the flow \cite{rightley97}. Since the particle characteristic size $d$ is usually small, so is the Reynolds number $Re=Ud/\nu$, where $U$ is the post-shock relative particle velocity and $\nu$ the kinematic viscosity of the embedding gas. Thus, in first approximation, the momentum exchange between the particle and the gas transpires via Stokes drag. However, a real shock-particle interaction is much more complex due to compressibility effects, resulting in Mach number dependence of the drag force \cite{bala1}. Transient forces in compressible flow can be much greater than the quasi-steady drag \cite{bala2}, leading to greater momentum exchange between 
the droplets (or particles) and the surrounding gas flow. With a sufficient particle concentration, this may result in the post-shock equilibrium local velocity of the gas-particle mixture being measurably lower than that of the unseeded gas. This entrainment of gas by multiple particles or droplets would be driven by viscosity, which dominates the flow on very small scales, somewhat akin to the recently observed entrainment of water by swarms of swimming plankton\cite{dabiri}.

\begin{figure}
\centerline{\includegraphics{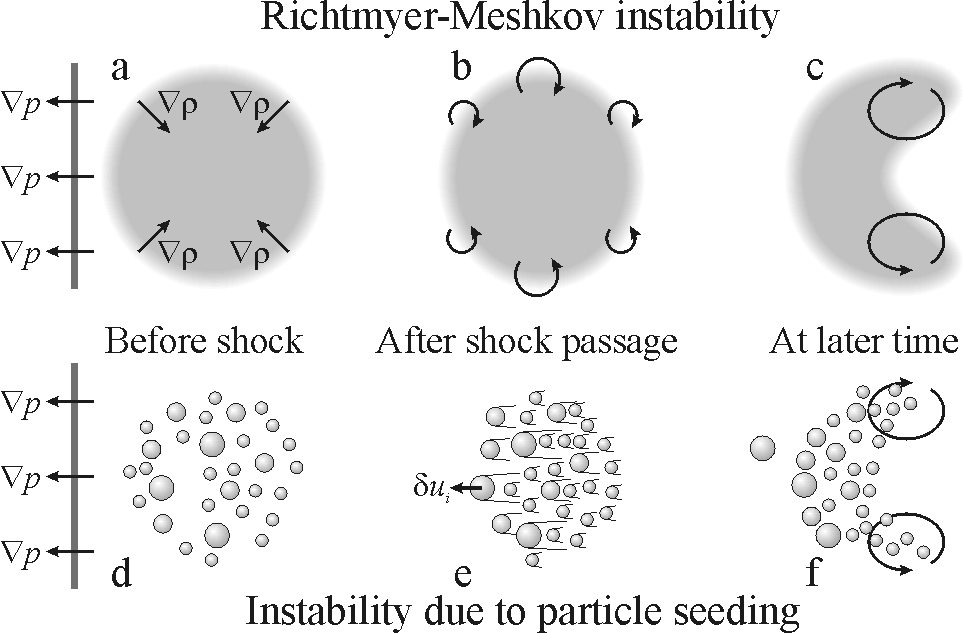}}
\caption{\label{explain} Schematic comparison of RMI (a-c, top row) and seeding-induced instability (d-f, bottom row). RMI is driven by misalignment between the
pressure gradient on the shock ($\nabla p$, a, d) and density gradient on the interface between two gases ($\nabla \rho$, a). This misalignment leads to vorticity deposition on the interface (b), and to subsequent vortex roll-up (c). In the seeded flow (d), there is no macroscopic density interface. Immediately after shock interaction, small particles lag behind the gas accelerated to piston velocity, acquiring relative velocities $\delta u_i$ (e) and interacting with the surrounding gas. As the particle and gas velocities reach equilibrium, the average velocity of the medium is lower than that of the unseeded gas, leading to shear and vortex roll-up (f).}
\end{figure}

Here we present experimental observations of what can be described as a general instability of a gas-droplet mixture with initially nonuniform 
droplet concentration developing after planar shock acceleration. While the morphology that emerges is superficially similar to that seen in
many RMI experiments \cite{RMI1,RMI2}, the physical mechanism responsible for the vortex formation in the flow is quite different (Fig.~\ref{explain}).

 The data presented in this Letter were acquired using a horizontal shock tube with a $75 \times 75$~mm cross-section. 
The driver section at the end of the shock tube is pressurized with helium. Subsequently, 
a planar shock wave is released into the driven section by puncturing the diaphragm 
initially containing pressurized helium in the driver section. The velocities and 
pressures associated with the shock wave are measured with two high-speed pressure 
transducers separated by a 2.59~m distance along the driven section. A horizontal 
cross-section of the flow is illuminated by a sequence of four laser pulses 
(each approximately 5~ns in duration) produced by frequency-doubled Nd:YAG lasers. 
Images of the flow are captured with a multiple-CCD (charge-coupled 
device) intensified IMACON 200 digital camera.

\begin{figure}
\centerline{\includegraphics[width=3.2in]{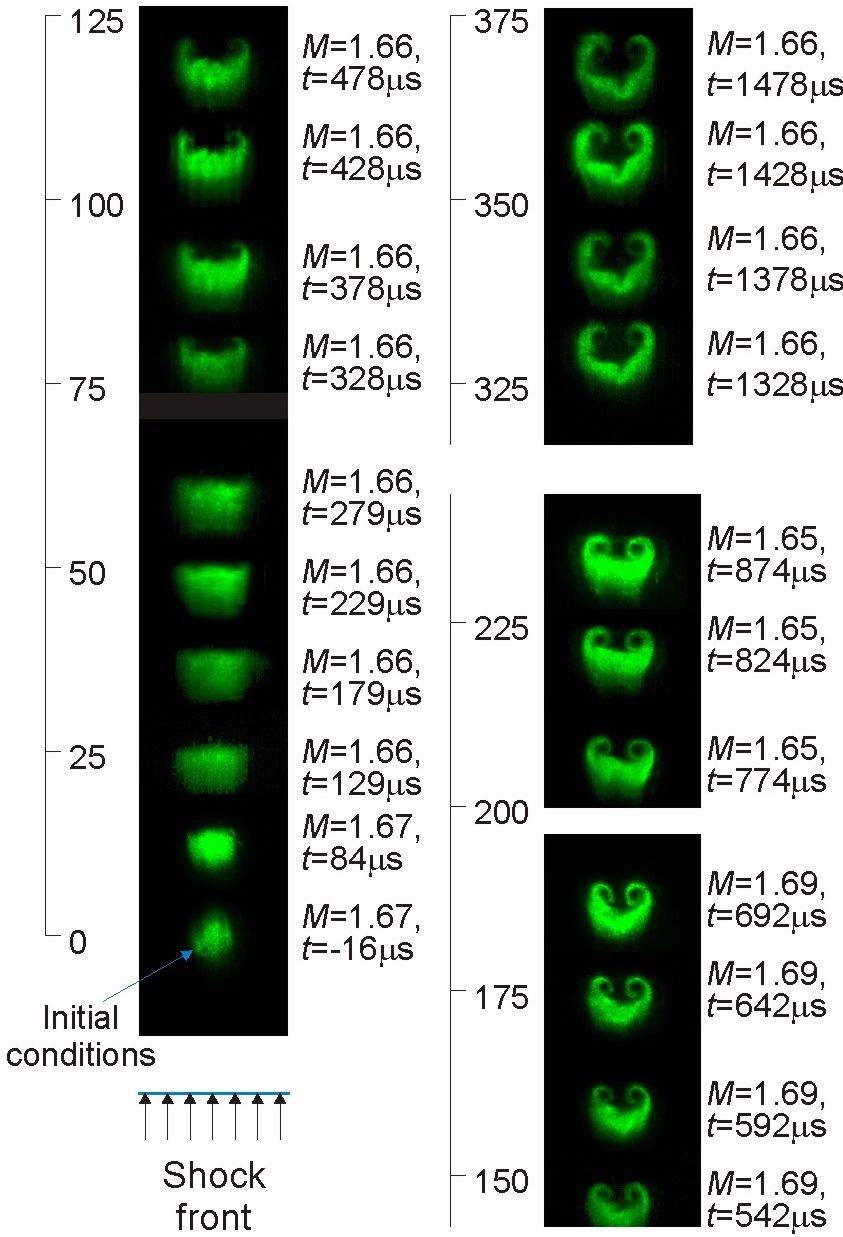}}
\caption{\label{fig1} Mosaic of six experimental image sequences showing the evolution 
of a shock-accelerated column of glycol droplets in air. Shock direction is from bottom 
to top. Timings of individual exposures (measured from shock impact on the droplet-seeded 
column) and Mach numbers corresponding to each image are marked in the figure, with time 
$t=0$ corresponding to the shock wave reaching the center of the initial conditions. 
Numbers on the left of the images denote distance from the center of the initial position 
of the column in mm.}
\end{figure}

In our experiments, the initial conditions are produced by vertically injecting a 
slow ($\sim 6$~cm/s) jet of air pre-mixed with micron-sized dipropylene glycol 
droplets into the optically transparent test section of the shock tube, filled 
with quiescent unseeded air. The particle size distribution in the jet was measured 
with a Malvern Spraytec particle sizer. This injection through 
a 6.35~mm diameter round nozzle 
forms a quasi-two-dimensional cylindrical column of seeded air. 
The image sequence in Fig.\ref{fig1} shows 
the evolution of the laser-illuminated cross-section of this column, as it and 
the surrounding air are accelerated by a planar shock moving at an average velocity 
of about 570~m/s (Mach number $M = 1.66 \pm 0.02$, with small variations from experiment to experiment). The shock 
passage accelerates the air behind it to a piston velocity of about 303 m/s. The glycol droplets are about 850 times 
more dense than air before shock compression and occupy a small ($\sim 5$\%) volume fraction of the injected column.

The initial conditions of the experiment are nearly two-dimensional, with little 
variation in the direction normal to the plane of view. In the earlier images of the 
sequence of Fig.~\ref{fig1} (third to fifth), the smallest droplets (submicron-sized) 
form a line parallel to the shock front, and larger droplets 
(average size 4~$\mu$m) trail farther upstream, forming a tail-like structure, 
because smaller droplets achieve momentum equilibrium with the embedding flow faster than the larger ones. Overall, this momentum exchange leads to the air-droplet mixture mean 
velocity in the section of the flow containing paerticles being lower than that of 
the surrounding particle-free air. 
Shear between the areas with lower and higher velocities in turn leads to formation 
of counter-rotating vortices on both sides of the column which are apparent in the seventh 
and subsequent images in Fig.~\ref{fig1}.

\begin{figure}
\centerline{\includegraphics{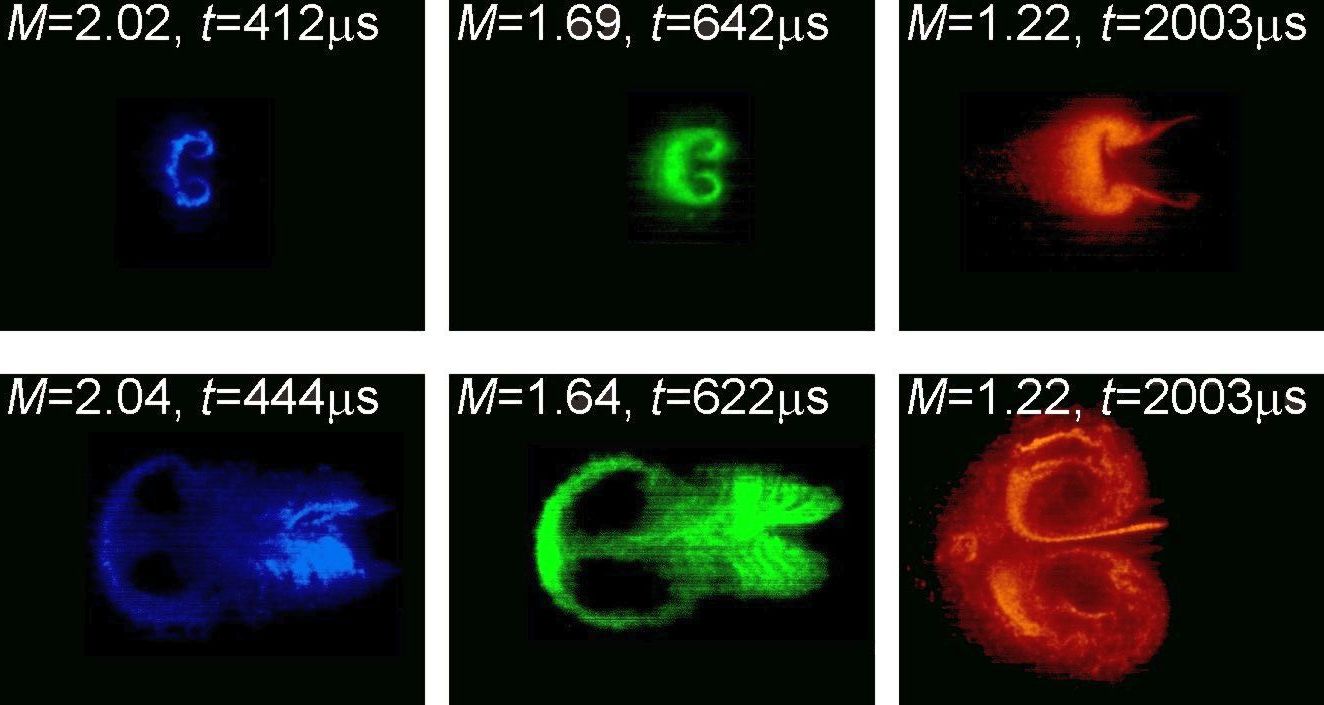}}
\caption{\label{fig2} Images of the instability of a droplet-seeded air column (top) and droplet-SF$_6$ column (bottom) at the same downstream position in the flow at different Mach numbers. Original images are illuminated with a green laser, false color is added here to distinguish between lower (red), intermediate (green) and high (blue) Mach numbers. The shock direction is from left to right, the image extent in the streamwise (horizontal) direction is from 152 mm to 181 mm downstream. Times from shock acceleration of the initial conditions and Mach numbers are labeled in each image.}
\end{figure}

The morphology that develops at later times as the result of this shock-driven interaction is superficially similar to that of a cylinder of heavy gas initially 
embedded in lighter gas and accelerated by a planar shock \cite{gani}, where two 
counter-rotating vortices form due to RMI. However, what happens in the two cases on the microscopic scale is quite different.
As there is no gaseous density interface, there is no baroclinic vorticity deposition. 
There are also discernible differences in the flow 
morphology due to RMI and to seeding-induced flow, as Fig.~\ref{fig2} illustrates. In this figure, we compare the evolution of the flow with droplet-seeded air column with that 
where the injected gas is sulfur hexafluoride SF$_6$, which is about five times more dense
than air. With SF$_6$ injection, an initially cylindrical, diffuse interface forms between the heavy gas and the lighter gas (air), resulting in a density gradient being added 
to the initial conditions, producing RMI upon shock passage. The images in 
Fig.~\ref{fig2} were acquired at the same downstream location, at three different 
Mach number for each case (RMI and seeding-induced flow). In the case of RMI, we still use droplet seeding, as many earlier experiments also did, 
for the purpose of flow visualization. Strictly speaking, this Letter demonstrates that the presence of these droplets may alter the flow, however, as 
the following discussion will show, in the presence of a shock-accelerated density interface, the flow is dominated by vortices developing from 
baroclinic vorticity deposition, and the vortex roll-up due to droplets alone is comparatively much weaker. 

The counter-rotating vortex pairs are evident in all the cases. However, for the case of RMI, another feature gains prominence as the Mach number increases, namely a central spike that emerges due to shock focusing in the heavy gas. This feature is notably absent in the images of the droplet-seeded column. RMI images also manifest small-scale structures that form due to secondary instabilities and lead to the flow transitioning to turbulence. In the case of RMI, the flow behavior at small scales associated with these features is known to become statistically consistent with models of fully developed turbulence \cite{gani}.

The small-scale structures visible in the right column of Fig.~\ref{fig2} arise both due to 
shear (Kelvin-Helmholtz instability) and to the secondary baroclinic instability 
induced by pressure-density misalignment in the cores of the vortices comprising the counter-rotating pair. Some evidence of the shear-driven secondary structures may be present in the late-time ($t>1$~ms) images of the droplet-column instability as well. However, even at the latest times observed, we did not see evidence of transition to turbulence for that case. Another notable feature of Fig.~\ref{fig2} is the rather weak dependence of the apparent size of the counter-rotating vortex pairs on the Mach number. 

\begin{figure}
\centerline{\includegraphics{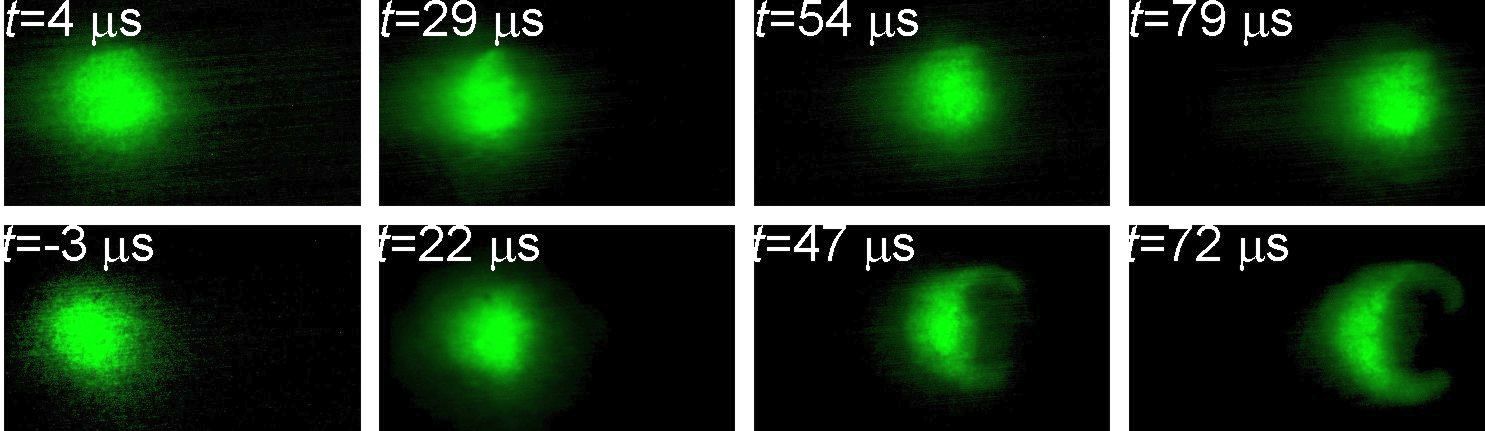}}
\caption{\label{fig3} Comparison of early-time evolution of air-droplet column (top) and SF6-droplet column (bottom) for $M~1.6$. Interval between frames is 25~ms, horizontal 
image extent is 24~mm, shock direction is from left to right.}
\end{figure}

In many earlier experiments, tracer particles or droplets were used under the assumption that they will behave as a passive scalar. This study demonstrates that such an assumption is generally not valid. However, in the presence of RMI, its much faster growth begins to dominate the flow from the time shortly after shock acceleration, as Fig.~\ref{fig3} illustrates. The subtle interplay of droplet or particle seeding with RMI will require further study. 

Despite the obvious differences in the formation of the instability of impulsively-accelerated two-phase
flow we describe here and in the formation of RMI in gas with a density interface, it is tempting to assess to which
extent the two shock-driven instabilities can be related. For RMI, the initial perturbation growth rate
according to Richtmyer's theory is $v^{imp}=-\Delta U A k a_0$, where $\Delta U$ is the difference between the mean velocity of the interface before and after the shock (in our case, the piston velocity with the negative sign), $a_0$ is the pre-shock amplitude of the initial interface perturbation, $k$ its characteristic wavelength, and $A$ is the Atwood number comprised of the densities of the gases on the upstream ( $\rho_1$) and downstream ( $\rho_2$) sides of the interface with respect to the direction of the shock: $A=(\rho_2-\rho_1)/(\rho_2+\rho_1)$. 
Realistically, this growth rate represents the initial slope to the curve describing the
actual interfacial growth, which is nonlinear. 
For the case of RMI on an air-SF$_6$ interface, $A \sim 0.67$.
For our experiment with droplets, we measured the average density of the droplet-seeded air ($1.26 \pm 0.01$~kg/m$^3$) and used that to compute the effective Atwood number $A \sim 0.03$.

\begin{figure}
\centerline{\includegraphics{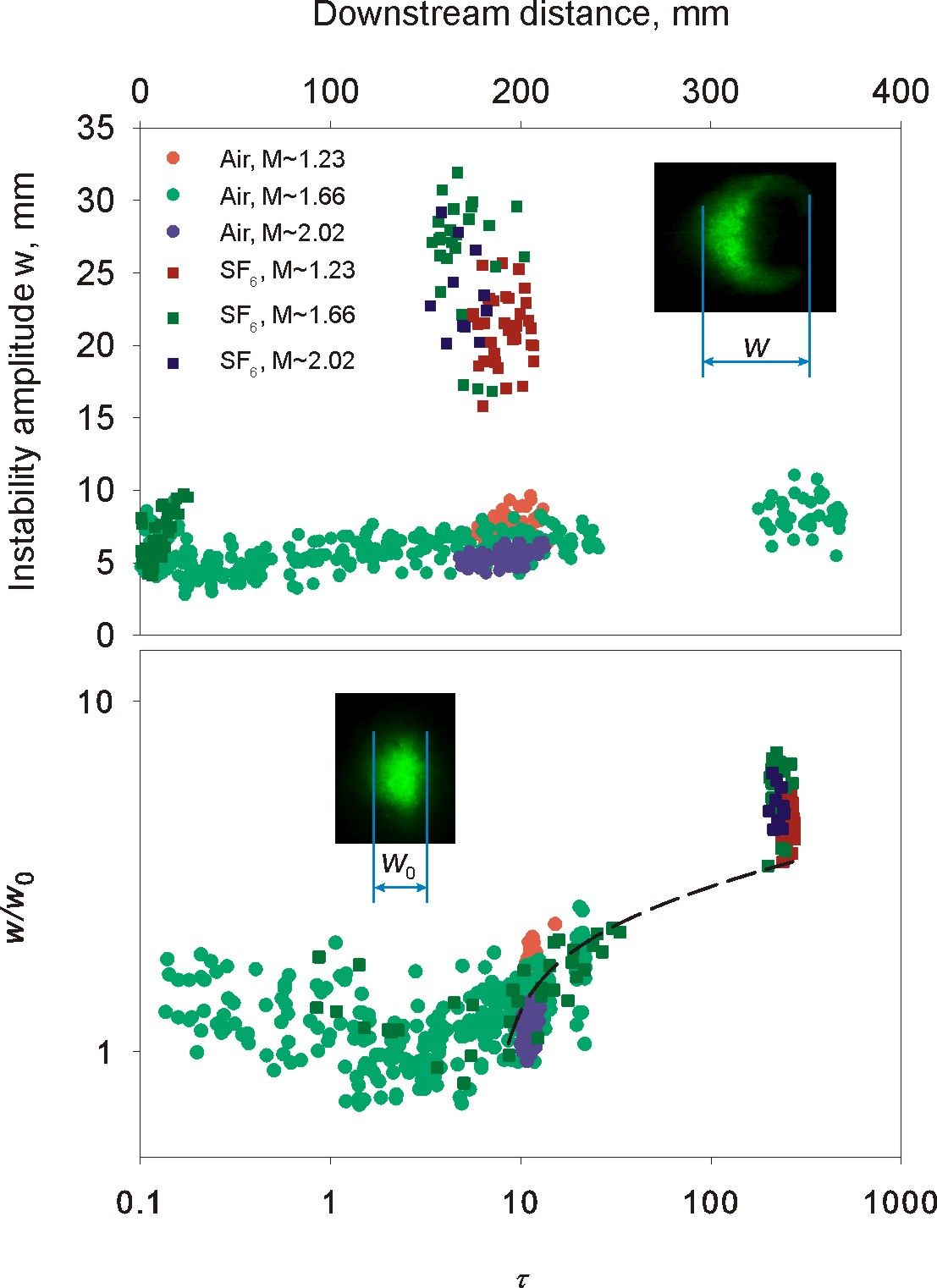}}
\caption{\label{fig4} Streamwise perturbation amplitude $w$ as the function of downstream distance in dimensional (top) and dimensionless (bottom) coordinates. Dashed line in 
the bottom plot represents a single-parameter fit to the experimental data for 
SF$_6$-droplet column at $M \sim 1.66$ using the modified Jacobs correlation 
\protect\cite{rightley97}. }
\end{figure}

The top plot of Fig.~{\ref{fig4}} shows a comparison of the instability growth for RMI and the instability of 
the droplet-seeded air column as the function of downstream distance from the initial conditions.
The data were acquired in about 200 experimental runs (resulting in 572 amplitude measurements), conducted in the range of Mach numbers from 1.22 to 2.02. In all cases, the amplitude initially decreases after the shock compression. Phase inversion effects on the downstream side of the gas cylinder are known to play a role in this decrease for RMI. Subsequently the perturbation amplitude begins to grow, much faster for the case of RMI, as the plot shows. It is also apparent that, for each case, the instability amplitude as the function of downstream distance indeed depends rather weakly on the Mach number.  

For RMI, it has been reported \cite{orlicz} that the instability amplitude growth 
curves produced from the same initial conditions at different Mach numbers can be 
collapsed by plotting them in appropriate dimensionless coordinates. The dimensionless 
temporal coordinate $\tau$ relates to time after shock acceleration $t$ as 
$\tau =2 k A |\Delta U| t$. In other words, for the same initial conditions and Atwood number,  plotting the RMI amplitude as the function of downstream distance $x= |\Delta U| t$ 
rather than time should also cause the growth curves for different Mach numbers to
collapse with appropriate scaling of the vertical axis. 

In the bottom plot of Fig.~\ref{fig4}, the horizontal coordinate is rescaled in 
terms of dimensionless time $\tau$, and the vertical coordinate is nondimensionalized
by $w_0$ – the average minimum width of the perturbed column after the shock 
acceleration for each subrange of Mach numbers ($M=1.22\pm 0.01$, $M=1.66 \pm 0.02$, 
$M=2.02 \pm 0.01$) –- to account for compression and phase inversion effects. 
Note that this non-dimensionalization should make the initial slopes of the instability 
growth according to Richtmyer's theory (after the compression and phase inversion effects 
took place) equal for all the Mach and Atwood numbers. 

Notably, non-dimensionalization produces a plot with fairly consistent growth trends both 
for RMI and for the instability of the two-phase droplet-seeded column. This may suggest 
that the long-term behavior of the shocked two-phase medium is consistent with that of 
continuous medium with the same average density. 

The dashed curve in the bottom plot is produced by using several flow parameters measured 
during the experiments, such as the post-shock minimal streamwise width of the perturbed 
column $w_0$ and the corresponding time $t_0$. These parameters are inserted into a semi-analytical 
formula \cite{rightley97} developed from an earlier formula \cite{jacobs}  that is known to faithfully describe nonlinear vortex
growth after shock acceleration, although with a somewhat different initial geometry \cite{rightley97,orlicz}:
$$
w(t)={\frac{2}{k}}{\rm sinh}^{-1}\left[ k^2 \Gamma \left(t - t_0 \right) + {\rm sinh} \left(\frac{k w_0}{2} \right) \right] 
$$
Here we can replace $t$ with $x/|\Delta U|$.
With this formula, we curve-fit our our experimental data for RMI at $M \sim 1.66$ 
using the shock-deposited circulation (that we do not measure directly) as the sole 
fit parameter. After the fit curve is rescaled consistently with the experimental 
data, it agrees reasonably well with both the measurements for SF$_6$ and the 
measurements for the air-droplet mixture. However, flow features present in RMI are 
absent in the case of the instability of the droplet-seeded column, namely the spike 
due to shock focusing and the secondary baroclinic instability in the cores of the counter-rotating vortices. Vice versa, the trailing tail of more massive 
particles characterizing the evolution of the air-droplet column is not prominent 
in RMI evolution. 

The instability described here can occur in a wide variety of impulsively accelerated two-phase flows of liquid, gas, and plasma with inclusions (e.g., after propagation of shocks through the cosmic dusty plasma). Moreover, it may be one of a class of similar instabilities – for example, non-uniform seeding of the top layer of a 
fluid or gas in a gravity field could produce an analog of Rayleigh-Taylor instability. 

\noindent \textbf{Acknowledgments.} We thank Professor S. ``Bala'' Balachandar (U. of Florida) for helpful suggestions. 
This research is funded by the US National Nuclear Security Agency (NNSA) under the Stewardship Science Academic Alliances program through DOE Grant DE-PS52-08NA28920 and by the US Defense Threat Reduction Agency (DTRA) under awards HDTRA1-07-1-0036 and  HDTRA1-08-1-0053.

\end{document}